\documentclass[twocolumn,showpacs,preprintnumbers,amsmath,amssymb,prb]{revtex4}

\usepackage{dcolumn}
\usepackage{graphicx}

\begin{document}
\bibliographystyle{apsrev}

\title{Microstructure dependence of low-temperature elastic properties \\in
amorphous diamond-like carbon films}

\author{Xiao Liu\footnote{Electronic mail: xiao.liu@nrl.navy.mil} and T. H. Metcalf}
\affiliation{Naval Research Laboratory, 4555 Overlook Ave. SW,
Washington DC, 20375-5320}
\author{P. Mosaner and A. Miotello}
\affiliation{Dipartimento di Fisica, Universit \`{a} di
Trento, 38050 Povo (TN), Italy}

\date{\today}

\begin{abstract}

We have studied the internal friction and the relative change in the speed of sound of amorphous diamond-like carbon films prepared by pulsed-laser deposition from 0.3\,K to room temperature. Like the most of amorphous solids, the internal friction below 10\,K exhibits a temperature independent plateau. The values of the internal friction plateau, however, are slightly below the universal ``glassy range'' where the internal frictions of almost all amorphous solids lie. Similar observations have been made in our earlier studies in the thin films of amorphous silicon and amorphous germanium, and the behavior could be well accounted for by the existence of the low-energy atomic tunneling states. In this work, we have varied the concentration of $sp^3$ versus $sp^2$ carbon atoms by increasing laser fluence from 1.5 to 30\,J/cm$^{2}$. Our results show that both the internal friction and the speed of sound have a nonmonotonic dependence on $sp^3/sp^2$ ratio with the values of the internal friction plateau varying between $6\times 10^{-5}$ and $1.1\times 10^{-4}$. We explain our findings as a result of a possible competition between the increase of atomic bonding and the increase of internal strain in the films, both of which are important in determining the tunneling states in amorphous solids. In contrast, no significant dependence of laser fluence is found in shear moduli of the films, which vary between 220 and 250\,GPa. The temperature dependence of the relative change in speed of sound, although it shows a similar nonmonotonic dependence on laser fluence as the internal friction, differs from those found in thin films of amorphous silicon and amorphous germanium, which we explain as having the same origin as the anomalous behavior recently observed in the speed of sound of thin nanocrystalline diamond films.

\end{abstract}

\maketitle

\section{Introduction}
\label{introduction}

The origin of the vibrational low-energy excitations in virtually all amorphous solids is still a mystery in condensed matter physics. The problem is not the lack of an experimental database or of theoretical descriptions, but rather the lack of a unified microscopic picture for the universality observed in almost all amorphous and certain disordered crystalline solids. Such microscopic picture would need to account for the remarkably similar behavior for a tremendously broad range of materials that differ in every aspect from microstructure to composition.\cite{02rmp74-pohl}

Below 1\,K, insulating amorphous solids have specific heat that is linear in temperature $T$ and thermal conductivity that varies as $T^2$, as was discovered in 1971 by Zeller and Pohl.\cite{71prb4-zeller} The internal friction of amorphous solids is constant in the so-called plateau region below approximately 10\,K to a characteristic temperature that depends on the measuring frequency, below which the internal friction starts to decrease and falls off eventually as $T^3$.\cite{76-mason-hunklinger} These and many other low-temperature properties of amorphous solids have been successfully explained by the phenomenological Tunneling Model, proposed by Anderson, Halperin, Varma,\cite{72pm25-anderson} and independently by Phillips,\cite{72jltp7-phillips} and extended by J\"{a}ckle,\cite{72zpb257-jackle} all in 1972. This model envisions that some atoms or groups of atoms sit in local double- or multi-well potentials inherent to their amorphous nature. At low enough temperatures, the individual wave functions overlap across the potential barrier, which results in atomic tunneling with little cost of energy, and this gives rise to a broad distribution of vibrational low-energy tunneling states, often called as two-level tunneling systems (TLS).

So far the most universal quantity of the Tunneling Model has been the so-called tunneling strength, defined as: 
\begin{equation} 
C=\frac{\bar P\gamma^2}{\rho v^2}, \label{c} 
\end{equation} 
where $\bar P$ is the spectral density of TLS, $\gamma$ their coupling energy to the lattice, $\rho$ the mass density, and $v$ the speed of sound. (Note that for simplicity, we have omitted the polarizations of phonon waves in this paper.) The quantity $C$ can be accessed experimentally either in the thermal conductivity below 1\,K or in the internal friction plateau below 10\,K, and $10^{-4}\leq C \leq 10^{-3}$ is found for almost all amorphous solids, even as the individual parameters, such as $\rho v^2$, can vary by as much as four orders of magnitude.\cite{03prb68-watson} In addition, $C$ is directly proportional to the ratio of the phonon wavelength $\lambda$ to the phonon mean free path $\ell$---a universal quantity that can be obtained from experiments without use of any theoretical model. The ratio $\lambda/\ell$ is found to lie between $10^{-3}$ and $10^{-2}$, even as $\lambda$ varies more than nine orders of magnitude in experiments.\cite{02rmp74-pohl}

In spite of its success, the Tunneling Model addresses neither the origin of TLS at the microscopic level nor the universal properties. Nevertheless, Phillips argued in his original work\cite{72jltp7-phillips} that TLS could occur naturally in some amorphous solids with an open structure, and are caused by atoms with a low coordination number that can tunnel among potential minima. He further speculated that if every atom were linked to more than two neighbors, the structure would be overconstrained and therefore rigid, so the double-well configurations would be unlikely. This should be the case for four-fold coordinated systems, like amorphous Si ({\em a}-Si) and amorphous Ge ({\em a}-Ge).

Since {\em a}-Si and {\em a}-Ge could only be prepared as thin films, early studies on {\em a}-Si and {\em a}-Ge were inconclusive due to the low sensitivity of the techniques used to measure thin films.\cite{80prl44-haumeder,85prl55-vandenberg,85pmb52-duquesne} Following those pioneering studies, we have used a high sensitivity technique to study the low-temperature internal friction of various types of thin films of {\em a}-Si, {\em a}-Ge, and also amorphous carbon ({\em a}-C).\cite{97prl78-liu,98prb58-liu,02pmb82-liu,04msea370-liu} We have concluded that the four-fold coordinated tetrahedral bonding is indeed an important factor in reducing the number of TLS.\cite{02pmb82-liu} In certain hydrogenated {\em a}-Si films, the TLS can be made to disappear completely.\cite{97prl78-liu} In some of the other cases, however, we have evidence to show that the overconstraint from tetrahedral bonding leads to an increase of internal strain. That adversely causes an increase of the density of TLS.\cite{04msea370-liu}

Alternatively, one might expect a progressive reduction of TLS with the increase of the mean coordination number $m$ in an amorphous system. According to the constraint-counting model proposed by Phillips\cite{79jncs34-phillips} in 1979 and later by Thorpe\cite{83jncs57-thorpe} in 1983, glasses can be divided into two categories: floppy and rigid. Rigidity percolates when the number of constraints per atom exceeds the number of degrees of freedom per atom. For a three-dimensional structure, the threshold is at $m=2.4$. It was speculated\cite{83jncs57-thorpe} that in the floppy region where $m<2.4$, the TLS may have their origin in the excess degrees of freedom. In surface acoustic measurements, Duquesne and Bellessa\cite{85pmb52-duquesne} observed in amorphous Se$_{1-x}$Ge$_{x}$ that $\bar P$ decreases by a factor of 4 from $x=0$ to $x=0.4$ ($2\leq m \leq 2.8$). But for $x=1$ (pure {\em a}-Ge, $m=4$), the $\bar P$ is the same as for $x=0.4$ ($m=2.8$), and no measurements were done for $0.4<x<1$ ($2.8<m<4$). Brand and v. L\"{o}hneysen\cite{91epl16-brand} measured the specific heat of amorphous As$_{x}$Se$_{1-x}$ and found a nonmonotonic dependence of $\bar P$ on $x$ after an initial decrease of $\bar P$ with increasing $x$ up to $x=0.4$ ($m=2.4$).

While it has been established that for $m<2.4$, $\bar P$ decreases with increasing $m$, studies are scarce on the rigid side of amorphous solids, where atoms are not only strongly bonded but also overconstrained. New knowledge of how atomic bonding and internal strain can interplay with each other and influence TLS may help us to understand the origin of the low-energy excitations in amorphous solids. 

Amorphous diamond-like carbon films\cite{remark1} have a unique combination of properties, such as high values of hardness, elastic moduli, electrical resistivity, and chemical inertness. These properties have stimulated considerable research and development interest for their applications in recent years; for a review, see Ref.\,\onlinecite{02mser37-roberts}. The keys that lead to films with these superior properties are the ability to prepare films with low internal strain and with a high concentration of $sp^3$ atoms relative to $sp^2$ atoms.\cite{97apl71-friedmann,00epl50-bonelli} In this work, we systematically study the low-temperature internal friction of amorphous diamond-like carbon films prepared by pulsed laser deposition (PLD). We vary the laser fluence so that we can control the $sp^3$ and $sp^2$ concentrations in the films to modify the mean coordination number between $m=3$ and 4. Our results suggest an interplay between atomic bonding and internal strain.

\section{Experimental}
\label{experimental}

\begin{table}
\caption{\label{tab:table1}Laser fluence, number of pulses, film thickness, room temperature shear modulus, and internal friction at 1\,K are listed below.}
\begin{ruledtabular}
\begin{tabular}{ccccc}
Fluences&Pulses&Thickness&$G_{\rm film}$&$Q_{0}^{-1}$ ($T=$1\,K)\\
(J/cm$^2$)& $\times 10^3$ &(nm)&(GPa)&$\times 10^{-4}$\\
\hline
30 & 20 & 142 & 226 & $0.86$ \\
26 & 22 & 155 & 223 & $1.00$ \\
18 & 25 & 200 & 244 & $1.32$ \\
10 & 25 & 221 & 254 & $1.42$ \\
5 & 18 & 156 & 220 & $1.19$ \\
1.5 & 50 & 209 & 252 & $0.80$ \\
\end{tabular}
\end{ruledtabular}
\end{table}

The films were deposited at the University of Trento at laser fluences ranging between 1.5 to 30\,J/cm$^2$, using highly oriented pyrolitic graphite as target.\cite{99mst10-bonelli} The laser source was a KrF excimer laser of 248\,nm wavelength, 20\,ns pulse duration, and 10\,Hz repetition rate. In these experiments the deposition chamber was evacuated to reach pressures lower than $1.5\times 10^{-6}$\,torr before laser irradiation. The target-to-substrate distance was fixed at 70\,mm. The film deposition parameters have been chosen in such a way that films would be deposited with appropriate atomic coordination, as carefully investigated in the Trento laboratory.\cite{00epl50-bonelli,02epjb25-bonelli,03jap93-bonelli} Table~\ref{tab:table1} lists some of the film deposition parameters. According to refs.\,\onlinecite{00epl50-bonelli} and \onlinecite{02epjb25-bonelli}, the $sp^3$ concentration would vary from 40 to 80\%, as measured on similarly prepared films.

\begin{figure} 
\includegraphics[width=3in]{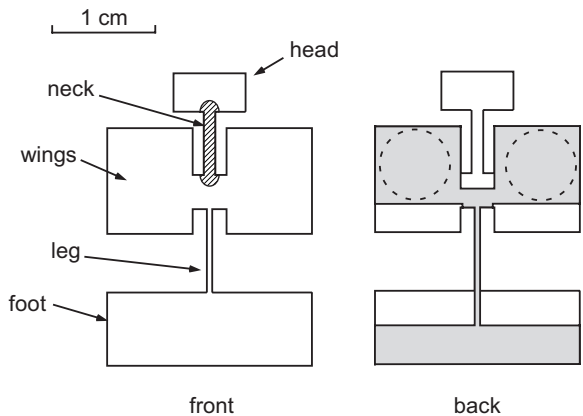} 
\caption{\label{fig1} Outline of the double-paddle oscillator. The left side shows the front view, and the right side shows the back view. The hatched area in the front view is the amorphous diamond-like carbon film deposited through a stainless steel mask. On the back side of the paddle a thin Cr and Au film is covered in the gray area. The dashed circles indicate the position of electrodes for capacitive drive and detection.} \end{figure}

Measurements of internal friction were performed using the double-paddle oscillator (DPO) technique.\cite{95mrs356-white} The DPOs were fabricated out of high purity P doped silicon wafers, which were $\langle 100 \rangle$ oriented and had resistivities $>5$\,k$\Omega$cm. The overall dimension of a DPO is 28\,mm high, 20\,mm wide, and 0.3\,mm thick; see Fig.\,\ref{fig1}. The DPO consists of a head, a neck, two wings, a leg, and a foot. The main axes are along the $\langle 110 \rangle$ orientation. On the back of the DPO a metal film (30\,{\AA} Cr and 500\,{\AA} Au) was deposited from the foot up to the wings but not on the neck and the head. The DPO was then clamped to an invar block using invar screws and a precision torque wrench. This minimized the effect of thermal contraction during cool down and ensured reproducibility after repeated remounting of the same DPO. Two electrodes were coupled to the wings from the back side so that the DPO could be driven and detected capacitively. For our internal friction measurements, we used the so-called second antisymmetric mode oscillating at $\sim$5500\,Hz. It has an exceptionally small background internal friction $Q^{-1}\approx 2\times 10^{-8}$ at low temperatures ($T<10$\,K) which is reproducible within $\pm 10$\% for different DPOs. The small $Q^{-1}$ is attributed to its unique design and mode shape. During oscillation, the head and the wings vibrate against each other, which leads to a torsional oscillation of the neck while leaving the leg and the foot with little vibration, minimizing the external loss. The film to be studied was deposited onto the neck (hatched area in Fig.\,\ref{fig1}). The internal friction results presented in this work were obtained exclusively using this mode for maximum detection sensitivity.

Deposition of a thin film onto the DPO changes its internal friction, $Q_{\rm osc}^{-1}$, as well as its resonance frequency, $f_{\rm osc}$, from those of a bare DPO, $Q_{\rm sub}^{-1}$ and $f_{\rm sub}$, respectively. From the difference, the shear modulus and the internal friction of the film can be calculated through
\begin{equation}
Q_{\rm film}^{-1} = \frac{G_{\rm sub} t_{\rm sub}}{3 G_{\rm film}
t_{\rm film}} (Q_{\rm osc}^{-1} - Q_{\rm sub}^{-1})+ Q_{\rm
osc}^{-1}, \label{Q-film}
\end{equation}
\begin{equation}
\frac{f_{\rm osc} - f_{\rm sub}}{f_{\rm sub}} = \frac{t_{\rm film}
}{2 t_{\rm sub}} \Bigl[\frac{3 G_{\rm film}}{G_{\rm sub}} -
\frac{\rho_{\rm film}}{\rho_{\rm sub}}(1+\eta)^{-1}\Bigr],
\label{f-film}
\end{equation}
where $t$, $\rho$, and $G$ are thicknesses, mass densities, and shear moduli of substrate and film, respectively; $\eta$ is the ratio of moments of inertia of the uncoated versus the coated part of head and neck, see Fig.\,\ref{fig1}. We include here a correction term in Eq.\,\ref{Q-film}, so that it will be valid even if $Q_{\rm film}^{-1}$ is comparable to $Q_{\rm sub}^{-1}$.\cite{04msea370-li} In Eq.\,\ref{f-film}, $\eta$ depends on the geometry and the coverage of the deposited film. For the film geometry shown in Fig.\,\ref{fig1}, we have $\eta=72$.  The shear modulus of the neck of the DPO along the $\langle 110 \rangle$ orientation is $G_{\rm sub}=6.2\times 10^{11}$\,dyn/cm$^{-2}$. In order to obtain $f_{\rm sub}$, we used CF$_4$ and O$_2$ based reactive ion etching to remove the {\em a}-C films from their substrates, followed by an oven anneal at 700 $^{\circ}$C for one hour to recover possible reactive ion damage and to remove carbon residuals. $f_{\rm sub}$ was then measured at room temperature. Since $f_{\rm sub}$ changes about 0.5\% from room temperature to 4\,K, we assume $G_{\rm sub}$ and $G_{\rm film}$ to be temperature independent in Eqs.\,\ref{Q-film} and \ref{f-film}.

\section{Elastic properties: the Tunneling Model}
\label{elastic}

The Tunneling Model explains and predicts a variety of low-temperature properties of amorphous solids. For comparison with experimental results in the later sections, we summarize here the expected elastic properties of insulating amorphous solids. More details can be found in Ref.\,\onlinecite{87rpp50-phillips}. 

The internal friction, which measures the elastic energy loss per oscillation cycle relative to the total elastic energy stored in the system, can be described in two limiting cases. At low enough temperatures, the maximum relaxation rate, $\tau^{-1}_{\rm max}$, of TLS is too slow to respond to the oscillations of the strain field at a frequency, $\omega$, i.e. $\tau^{-1}_{\rm max}\ll \omega$. As a result, the contribution of TLS to internal friction drops off as $T^3$. At higher temperatures, where $\tau^{-1}_{\rm max}\gg \omega$, there always exist some TLS with a relaxation rate $\tau^{-1}$ that satisfies $\omega \tau=1$ to cause a typical Debye relaxation event. Together with the fact that the density of TLS is constant in both energy and relaxation time---a fundamental assumption of the Tunneling Model---one gets a temperature independent internal friction: 
\begin{eqnarray}
Q^{-1}& = & \frac{\pi^4}{96}C\Bigl[\frac{T}{T_{\rm co}}\Bigr]^{3} \quad(T\ll T_{\rm co}), \label{q-low}\\
Q^{-1} & = & Q_{0}^{-1}=\frac{\pi}{2}C \quad(T\gg T_{\rm co}), \label{q-high}
\end{eqnarray}
where $T_{\rm co}$ is the crossover temperature determined by $\omega \tau^{-1}_{\rm max}=1$. We call $Q_{0}^{-1}$ the internal friction plateau. Since $Q_{0}^{-1}$ is directly proportional to the tunneling strength $C$ defined in Eq.\,\ref{c}, the internal friction plateau is not only independent of temperature but also independent of the measuring frequency. It should be as universal as $C$, varying with materials being measured between $1.5\times 10^{-4}$ and $1.5\times 10^{-3}$.

The same relaxational process that contributes to the internal friction also contributes to the speed of sound at $T\gg T_{\rm co}$ as
\begin{equation}\label{dv-rel}
\frac{\Delta v}{v_0}\Big|_{\rm rel}=-\frac{3}{2}C{\rm ln}\frac{T}{T_0},
\end{equation}
where $v_0$ is the speed of sound at an arbitrary reference temperature $T_0$, typically the lowest temperature being measured. There is also a resonant contribution to the speed of sound at any given temperature:
\begin{equation}\label{dv-res}
\frac{\Delta v}{v_0}\Big|_{\rm res}=C{\rm ln}\frac{T}{T_0}.
\end{equation}
Combining both, we have:
\begin{eqnarray}
\frac{\Delta v}{v_0} & = & C{\rm ln}\frac{T}{T_0} \quad(T\ll T_{\rm co}),  \label{v-low}\\
\frac{\Delta v}{v_0} & = & -\frac{1}{2}C{\rm ln}\frac{T}{T_0} \quad(T\gg T_{\rm co}). \label{v-high}
\end{eqnarray}
This results in a maximum in $\Delta v/v_0$ {\em vs.} $T$ at $T_{\rm co}$.

As the temperature goes higher to above a few Kelvin and up to tens of Kelvin, Bellessa\cite{78prl40-bellessa} found that the $\Delta v/v_0$  deviates from the logarithmic temperature dependence given in Eq.\,\ref{v-high} and varies linearly with temperature:
\begin{equation}
\frac{\Delta v}{v_0}= -\beta(T-T_0). \label{delta-v}
\end{equation}
By compiling data on the linear variation of the speed of sound described by Eq.\,\ref{delta-v}, White and Pohl\cite{96zpb100-white} summarized an empirical relationship between the slope, $\beta$, and the internal friction plateau $Q_0^{-1}$:
\begin{equation}
\beta =\frac{1}{2} Q^{-1}_{0} ({\rm K}^{-1}).
\label{eq:beta}
\end{equation}
This relation was found to hold for both amorphous solids and disordered crystals that behave like a glass.\cite{96zpb100-white} It was later found to hold for {\em a}-Si and {\em a}-Ge as well.\cite{98prb58-liu}

\section{Results}
\label{results}

\begin{figure} 
\includegraphics[width=3in]{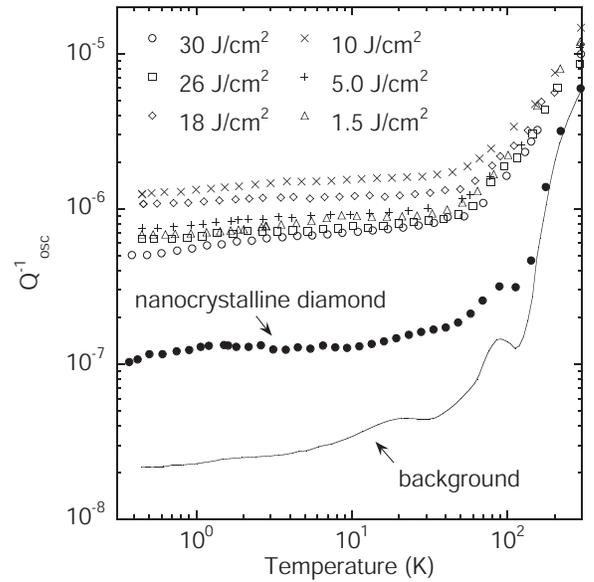} 
\caption{\label{fig2} The internal frictions of double-paddle oscillators carrying {\em a}-C films with different laser fluences labelled in the figure, and a 350\,nm thick nanocrystalline diamond film taken from Ref.\,\onlinecite{metcalf}. The background internal friction of a bare double-paddle oscillator is shown as a solid line.} 
\end{figure} 
\begin{figure} 
\includegraphics[width=3in]{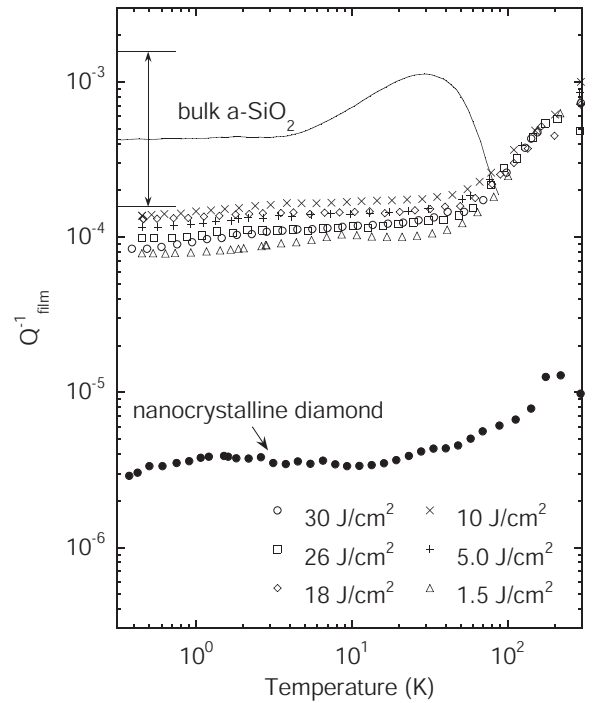} 
\caption{\label{fig3} The internal frictions of {\em a}-C films with different laser fluences labelled in the figure, and of a 350\,nm thick nanocrystalline diamond film taken from Ref.\,\onlinecite{metcalf}. The internal friction of bulk a-SiO$_2$, taken from Ref.\,\onlinecite{vancleve91}, is shown for comparison. The double arrow denotes the "glassy range" explained in the text.} 
\end{figure} 
The internal frictions of the six DPOs carrying {\em a}-C films are shown in Fig.\,\ref{fig2}, together with a DPO carrying a nanocrystalline diamond film deposited by chemical vapor deposition.\cite{metcalf} The solid line labelled ``background" is the internal friction of a bare DPO, $Q_{\rm sub}^{-1}$, used in Eq.\,\ref{Q-film}. To calculate $Q_{\rm film}^{-1}$, we need to obtain $G_{\rm film}$ from Eq.\,\ref{f-film}. Since the second term in Eq.\,\ref{f-film} is about two orders of magnitude smaller than the first, we drop it from our calculation. The results are shown in Table~\ref{tab:table1}. We find that $G_{\rm film}$ varies between 220 and 254\,GPa and does not depend on laser fluence or $sp^3/sp^2$ ratio in any significant way. For convenience, we use $G_{\rm film}=250$\,GPa in our calculation for all six films. Although $sp^3/sp^2$ ratio dependence has been reported for elastic moduli, the shear moduli determined in this work are within the range of the literature values: Using a surface Brillouin scattering technique, Ferrari {\em et al.}\cite{99apl75-ferrari} determined the shear modulus $G$ of a 76\,nm thick {\em a}-C film with 88\% $sp^3$ deposited by an S-bend filtered cathodic vacuum arc to be 337\,GPa. Using the same deposition and measuring techniques, Beghi {\em et al.}\cite{02msec19-beghi} later reported $G=130$--210\,GPa for an 8\,nm thick film and $G=70$--130\,GPa for a 4.5\,nm thick film. Marques {\em et al.}\cite{03apl83-marques} reported Young's modulus could increase from 40 to 120\,GPa in a series of hydrogenated {\em a}-C films when $sp^3$ concentration increases from 40 to 65\%. 

Using Eq.\,\ref{Q-film}, we show in Fig.\,\ref{fig3} the internal frictions of the six {\em a}-C films as well as the nanocrystalline diamond film for comparison. Both the {\em a}-C films and the nanocrystalline diamond film have an internal friction quite independent of temperature below 20\,K. The {\em a}-C films have at least one order of magnitude higher internal frictions than that of the nanocrystalline diamond film, although one would expect the internal friction of nanocrystalline diamond film to be much smaller. The internal friction of the nanocrystalline diamond film is attributed to a highly disordered interfacial layer between the film and silicon substrate, where the internal friction could be even higher than in its amorphous counterpart.\cite{metcalf} Thus, for the most part of the thickness above the interfacial layer, we could infer that the internal friction may indeed be much smaller. 

The internal frictions of the {\em a}-C films bear the signature of TLS in terms of its temperature independent plateau, $Q_{0}^{-1}$, at low temperatures, although its magnitude is slightly lower than in a typical amorphous solid, as noted in section \ref{elastic}. Apparently, the lowest measured temperature is not low enough to allow us to observe the $T^3$ drop-off of Eq.\,\ref{q-low}. In Fig.\,\ref{fig3}, the double arrow shows the range of the universal internal friction plateau, in which virtually all other amorphous solids lie. It is often called the ``glassy range." Also shown in Fig.\,\ref{fig3} is the internal friction of the prototypical glass, amorphous SiO$_2$ ({\em a}-SiO$_2$) measured at 4500\,Hz.\cite{vancleve91} In our previous study on {\em a}-C films prepared at the Naval Research Laboratory,\cite{04msea370-liu} we found that their internal friction plateaus can vary over an order of magnitude depending on laser fluences, dopings, and annealing, with the highest internal friction still below the glassy range. Our current results are consistent with those of the earlier study in the sense that the values of internal friction plateau below 20\,K fall in the same order of magnitude and they are all below the glassy range. In this study, we vary the laser fluence over such a wide range that we can observe a clear nonmonotonic variation of the internal friction. As the temperature increases, the internal frictions start to rise at about 50\,K, above which they show much less laser fluence dependence than at lower temperatures, similar to what we observed previously on unannealed films. 

\begin{figure} 
\includegraphics[width=3in]{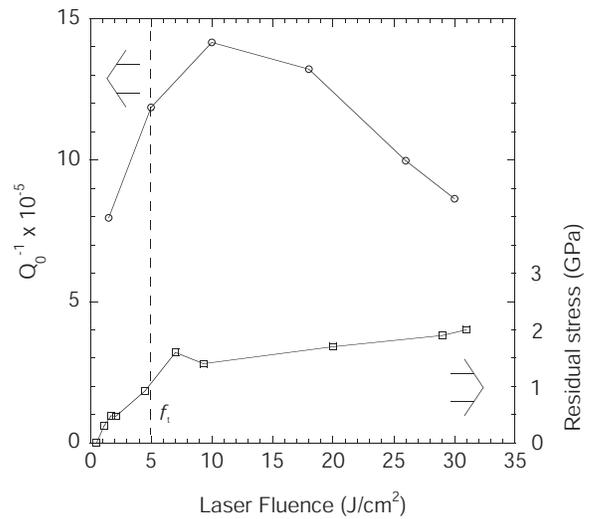} \caption{\label{fig4} The internal frictions of {\em a}-C films at 1\,K representing the plateau $Q_{0}^{-1}$ (left Y-axis) and residual compressive stress in the films after deposition (right Y-axis) versus laser fluence. The residual stress data are taken from Ref.\,\onlinecite{02epjb25-bonelli} measured on a series of similarly prepared {\em a}-C films. The vertical dashed line indicates the structural transition at $f_{\rm t}=5$\,J/cm$^2$ from primarily disordered graphitic glass to a tetrahedrally bonded amorphous solid.} 
\end{figure}

To illustrate the laser fluence dependence, we show in Fig.\,\ref{fig4} the internal frictions of {\em a}-C films at 1\,K (also listed in Table~\ref{tab:table1}), representing the plateau, versus laser fluence. 
The internal friction rises by a factor of 2 with the initial increase of laser fluence from 1.5\,J/cm$^2$ and reaches a maximum at 10\,J/cm$^2$. As the laser fluence further increases, $Q_{0}^{-1}$ drops off and becomes as low at highest fluence as it was at the lowest fluence. The variation of internal residual stress measured on a series of similarly prepared films by the same Trento group\cite{02epjb25-bonelli} is also plotted in Fig.\,\ref{fig4}. The stress is compressive. One can see that the initial increase of the internal friction correlates with the initial increase of the residual stress. At a laser fluence of about 10\,J/cm$^2$, the residual stress has reached saturation at about 1.6--2.0\,GPa which coincides with the maximum $Q_{0}^{-1}$. The vertical dashed line at 5\,J/cm$^2$ in Fig.\,\ref{fig4} indicates the structural transition of the {\em a}-C films from primarily disordered graphitic glass to a tetrahedrally bonded amorphous solid, as observed by the studies of Raman spectroscopy, infrared transmission spectroscopy, and X-ray reflectivity.\cite{00epl50-bonelli,02epjb25-bonelli} 

\begin{figure} 
\includegraphics[width=3in]{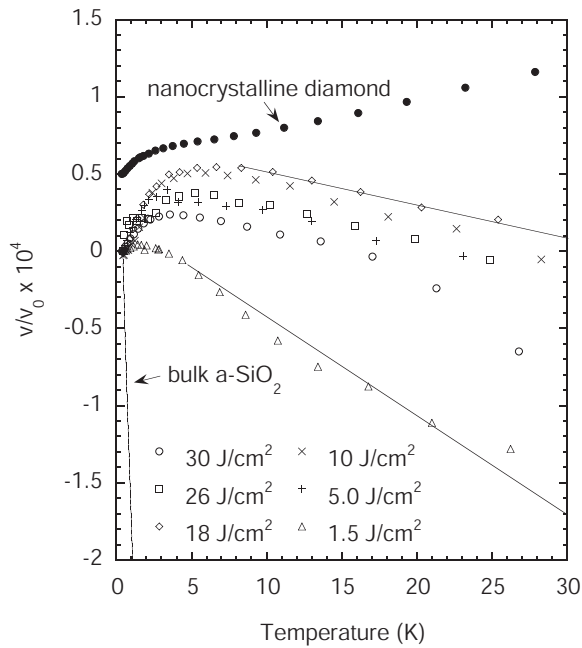} 
\caption{\label{fig5} The relative changes of the speed of sound for the six {\em a}-C films with different laser fluences labelled in the figure. The data for a 350\,nm thick nanocrystalline diamond film, taken from Ref.\,\onlinecite{metcalf}, are shown for comparison. It is shifted upward for clarity purpose. The data for bulk a-SiO$_2$, taken from Ref.\,\onlinecite{vancleve91}, are also shown for comparison. The two straight lines indicate the linear temperature dependence discussed in the text.} 
\end{figure}

Fig.\,\ref{fig5} shows the relative change of the speed of sound up to 30\,K for the six {\em a}-C films and the nanocrystalline diamond film. The data for the nanocrystalline diamond film has been shifted upwards for clarity. 
To our surprise, in this temperature range, $\Delta v/v_0$ for all {\em a}-C films does not follow the linear temperature dependence expressed in Eq.\,(\ref{delta-v}) as a typical amorphous solid. Instead, it increases with temperature, develops a maximum between 3 and 7\,K, and finally decreases with temperature linearly with a slope $\beta$ (Eq.\,\ref{delta-v}) varying between $2.1\times 10^{-6}$ and $5.7\times 10^{-6}$\,K$^{-1}$, more than 30 times smaller than that of the bulk {\em a}-SiO$_2$, also shown in Fig.\,\ref{fig5}. The location of the maxima depends strongly on laser fluence. In fact, both the height and the value of the maximum follow the same nonmonotonic dependence on laser fluence as the internal friction in the same temperature range. For crystals, one would expect a temperature independent $\Delta v/v_0$ in this temperature range. But the $\Delta v/v_0$ of the nanocrystalline diamond film shows an initial strong increase below 2\,K, superimposed by a slower increase that eventually develops a broad maximum at above 100\,K (not shown, see Ref.\,\onlinecite{metcalf} for details). The anomalous $\Delta v/v_0$ for both the {\em a}-C films and the nanocrystalline diamond film is exceptional and will be discussed below.

\section{Discussions}
\label{discussions}
\subsection{Tunneling states in {\em a}-C films}
\label{tunneling}

Taking into account of our previous studies,\cite{02pmb82-liu,04msea370-liu} we summarize that the internal friction plateaus of all {\em a}-C films that we have studied so far are below the ``glassy range." The smallest internal friction that we have ever observed, {\em a}-C doped with nitrogen,\cite{04msea370-liu} is about one order of magnitude below the ``glassy range." In {\em a}-Si films, we have reached an internal friction about three orders of magnitude below the ``glassy range" using hot-wire chemical vapor deposition that incorporates approximately 1 at.\% hydrogen.\cite{97prl78-liu} We believe that by exploring different deposition techniques and the parameter space, one could further lower the internal friction in {\em a}-C films. In contrast, among all the other amorphous solids studied so far, the lowest internal frictions that have been reported are only about a factor of 2 below the ``glassy range" in a couple of metallic glasses (for a review, see Ref.\,\onlinecite{02rmp74-pohl}). Hence, our current study gives further support to the conclusion that we reached earlier, which is the density of TLS is generally smaller in amorphous structures in which the atoms are predominately four-fold tetrahedrally bonded.

The main focus of this work is, however, to vary the laser fluence so that the mean coordination number could be varied systematically. The nonmonotonic dependence of $Q_0^{-1}$ on laser fluence in Fig.\,\ref{fig4} shows that the tunneling strength is more sensitive to the internal strain than to the mean coordination number. Nonetheless, as the internal strain reaches a saturation at a fluence of 10\,J/cm$^2$, we start to observe a clear dependence of $Q_{0}^{-1}$ with laser fluence: $Q_{0}^{-1}$ decreases with the increase of the fluence. This is consistent with Phillips' original argument\cite{72jltp7-phillips} and the constraint counting model\cite{79jncs34-phillips,83jncs57-thorpe} explained in the section \ref{introduction}. To our knowledge, this is the first experiment to study the coordination number dependence of TLS at $3<m<4$. In addition, our system, consisting of monatomic carbon atoms with either $sp^2$ or $sp^3$ bonding, does not suffer from a stoichiometric effect which is typical in binary or ternary systems.\cite{89prb39-tanaka} 

The important role played by internal strain in the generation of TLS in amorphous solids has been demonstrated before: Watson\cite{95prl75-watson} has shown that when less than 1 mole \% CN$^{-}$ ions are doped into either KCl or KBr crystal, CN$^{-}$ can cause Schottky anomalies by tunneling between degenerate orientation sites. This results in a narrow peak in specific heat, thermal conductivity, and internal friction at low temperatures. When the crystal is strained by intermixing KCl and KBr half and half, however, a broad distribution of tunneling states, as in an amorphous solid, occurs. One of our previous studies \cite{98prl81-liu} has shown that ion implantation into silicon crystals not only generates individual defects, such as divacancies, but also accumulates compressive strain by injection of energetic ions. An internal friction plateau as in an amorphous solid results as the strain saturates regardless of whether the material becomes amorphous or not. In this work, internal strain is caused by the injection of carbon ions into the film. The higher the laser fluence, the higher the internal strain, and the higher the internal friction plateau, until a saturation is reached. Our earlier study reached a similar, but less obvious, conclusion that the reduction of internal strain by means of doping and annealing results in a reduction of $Q_{0}^{-1}$ in {\em a}-C films.\cite{04msea370-liu} 

The competition between internal strain and tetrahedral bonding observed here helps to explain the question that if the TLS indeed originate from the excess degrees of freedom, why there is still a significant density of TLS left in {\em a}-C films which are predominately fourfold or threefold bonded. Watson\cite{95prl75-watson} demonstrated that as long as the internal strain is strong, a doping of CN$^{-}$ as small as 0.2 mole \% is enough to cause KCl-KBr crystal to behave like a glass. But the internal strain alone by intermixing KCl and KBr does not result in TLS. The situation is similar to our {\em a}-C films: the existence of individual local tunneling defects, no matter how small, is a necessity while the internal strain is vital in causing TLS and in determining the density and strength of the TLS. Had the residual stress not saturated in Fig.\,\ref{fig4}, $Q_{0}^{-1}$ might increase further. In other words, even though the density of local tunneling defects might be low in a primarily fourfold and threefold coordinated system, the TLS could be just as strong as in other amorphous solids in the presence of high internal strain. The same is true in {\em a}-Si: the smallest internal friction---three orders of magnitude below the ``glassy range"---is observed in films whose residual stress is essentially zero.\cite{98prb58-liu}

While the internal frictions of all the six {\em a}-C films shown in Fig.\,\ref{fig3} are typical for an amorphous solid, the maxima in $\Delta v/v_0$ at about 3--7\,K shown in Fig.\,\ref{fig5} may have a different origin. If the maxima were as a result of a transition from resonant to relaxational processes of TLS expressed in Eqs.\,\ref{v-low} and \ref{v-high}, one would expect the internal frictions in Fig.\,\ref{fig3} to start to drop off at about the same temperature as defined in Eqs.\,\ref{q-low} and \ref{q-high}. This is certainly not the case. Since the lower temperature sides of the maxima resemble that of the nanocrystalline diamond film also plotted in Fig.\,\ref{fig5}, we suggest that they may have a common origin which we still do not understand, such as an interfacial effect or a defect mode. 

For the nanocrystalline diamond film, the broad maximum in $\Delta v/v_0$ at about 100\,K has been attributed to the differential contraction between silicon substrate and nanocrystalline diamond film.\cite{metcalf} The same differential contraction effect is expected to be larger between an {\em a}-C film and its crystalline silicon substrate since the coefficient of linear thermal expansion of {\em a}-C films is about 2--3 times larger than in its crystalline counterpart.\cite{93jvsta11-blech,99jap86-lima} It is, therefore, reasonable to think that the negative slope of linear temperature variation above the maximum in $\Delta v/v_0$ expressed in Eq.\,\ref{delta-v} is superimposed onto a positive slope due to differential contraction. That could explain why $\beta$ is more than one order of magnitude smaller than $Q^{-1}_{0}/2$ as would be expected from Eq.\,\ref{eq:beta}.

\subsection{Microscopic origin and theoretical considerations}
\label{microscopic}

In order to gain some insight as to what has become of the TLS in {\em a}-C films, it is necessary to take a close look at the microstructure as it varies with laser fluence. Studies\cite{00epl50-bonelli,02epjb25-bonelli} have shown that the $sp^3/sp^2$ ratio does not follow the fluence in a linear fashion. Rather, a significant portion of the increase of the $sp^3$ concentration happens as the structure undergoes a transition from primarily disordered graphitic glass to a tetrahedrally bonded amorphous solid at a threshold fluence of 5\,J/cm$^2$. Starting at low fluences, $sp^2$ rings make a sizable proportion of $sp^2$ bonds. With increasing fluence, the concentration of $sp^2$ rings drops down quickly, vanishes even well before the threshold is reached, and is replaced by $sp^2$ chains and clusters. After passing the threshold fluence, the $sp^3$ concentration rises rapidly to about 80\%, and barely increases after that. In the middle range of fluences, $sp^3$ intermixes with $sp^2$ by breaking down $sp^2$ chains and clusters. A film has maximum disorder in this middle range, because there are a wide range of configurations with different local microstructure and phonon modes during the intermixing phase. Intermixing $sp^2$ and $sp^3$ also serves to release the internal strain that originates from the rigid tetrahedral bonding. 

The microstructure of the {\em a}-C films can also be viewed from a film growth point of view. Under pulsed laser flash, the ablated particles in the plume consist of C, C$_2$, C$^+$, and C$^{++}$. The film will be graphitic if only neutral, low-energy particles arrive on the substrate. The C$^+$ ions are directly responsible for the growth of tetrahedral structure. The higher the laser fluence, the higher energy of C$^+$ ions impinging onto the film; the excess energy densifies the film, generates compressive strain, and transforms $sp^2$ to $sp^3$. Depending on deposition conditions, high strain could be partially released by a self-annealing process. This situation is similar to the generation of TLS induced by ion-implantation in crystalline silicon.\cite{98prl81-liu} 

The mass density of our {\em a}-C films increases with laser fluence:\cite{02epjb25-bonelli} below the threshold, it is about 2.0--2.2\,g/cm$^3$; above that, 2.9--3.0\,g/cm$^3$. The nonmonotonic dependence of $Q_0^{-1}$ on fluence that we observed here, however, does not support a simple relationship between $Q_0^{-1}$ and the mass density. If we have to correlate the nonmonotonic dependence of $Q_{0}^{-1}$ in Fig.\,\ref{fig4} to its microstructure, it is the overall disorder caused by intermixing of $sp^2$ and $sp^3$ that $Q_{0}^{-1}$ follows, rather than any individual bonding configuration. Similar to {\em a}-Si and {\em a}-Ge films,\cite{98prb58-liu} {\em a}-C films are prepared under nonequilibrium conditions; local floppy modes may arise from locally defective, effectively underconstrained, regions due to inhomogeneity and internal strain of the films. More detailed microscopic mechanisms of TLS in tetrahedral systems have been suggested.\cite{99ep48-dyrting,03cmcc48-rivier} 

Based on our observations, the internal strain may have two different effects in generating TLS and in preserving the universality: The first is to spread the tunneling frequencies by modifying local potentials. The strain induced deformation may also remove some of the local tunneling defects, and the ultimate yield strain may keep the number of effective tunneling defects down to a certain level in an amorphous solid. This explanation is consistent with the classical noninteracting Tunneling Model. The second is to mediate strain interactions among the local tunneling defects that not only give rise to a continuous energy distribution but also limit the total density of TLS, as first proposed by Klein {\em et al.} in 1978.\cite{78prb15-klein}. The latter scenario seems to be gaining ground as more and more evidence of interactions of tunneling defects has been reported in recent years.\cite{burin98,98prl80-strehlow,00prl84-classen,02prb65-konig,03prl90-ladieu} In the present work, we are able to vary $C$ systematically by a factor of 2 in a relatively simple monatomic system; though it is not much, it may provide a unique testing ground for theories that tackle the problem of universality. Let's consider a few examples here: 

Based on the soft-potential model, Parshin\cite{94pss36-parshin} considered the interaction of all defects: not only the tunneling defects, but also defects capable of creating a deformation around themselves. He was able to explain the universality and the smallness of $C$, and to get an approximate correlation between $C$ and the macroscopic parameters of amorphous solids as $C\sim \rho^{1/3}v{\bar M}^{-5/6}M^{-1/2}$, where $\bar M$ and $M$ are the mean mass of the system and the effective mass of the tunneling object, respectively. For our {\em a}-C films, however, a variation of $C$ by a factor of 2 would require $M$ to vary by a factor of 4. Adopting a spin-glass approach, K\"{u}hn\cite{03epl62-kuhn} correlates $C$ to the glass transition temperature $T_{\rm c}$ as $C\sim T_{\rm c}^{-3/4}$, and a nonmonotonic variation of $T_{\rm c}$ on laser fluence would become a key test of the theory. Turlakov\cite{04prl93-turlakov} recently developed a elastohydrodynamic theory to explain the universality of $\lambda/l$, calculating $C\sim \lambda/l\sim (a/\xi)^3$, where $a$ and $\xi$ are interatomic distance and correlation length of disorder. This would indicate that $C$ is related to some kind of medium range order of an amorphous system. At this point, we can only say that more specific work is needed to understand the universality of $C$, and the parameters that could eventually tune it.  

\section{Conclusion}
\label{conclusion}

Amorphous carbon films, like {\em a}-Si and {\em a}-Ge films, have vibrational low-energy excitations that can be described by the Tunneling Model. Their density of TLS, however, is consistently smaller than most of the other amorphous solids, presumably due to their four-fold tetrahedral bonding. Varying the $sp^3/sp^2$ ratio from 40 to 80\% in the film leads to a nonmonotonic change of the internal friction plateau $Q^{-1}_{0}$ and, therefore, the tunneling strength $C$. This has been viewed as a competition between the effect of tetrahedral bonding and the internal strain. Our results emphasize the effects of random stress in the generation of TLS: as long as there is at least a minuscule amount of local tunneling defects, random stress will transform them into TLS, by either redistributing their energies, mediating mutual interactions, or both. The TLS in {\em a}-C films does not seem to correlate to any particular type of microstructure, rather, the overall disorder. The anomalous behavior found in the relative variation of speed of sound in {\em a}-C is attributed to the effects of substrate and interfaces that would occur in their crystalline counterparts as well. We hope the laser fluence dependence of $C$ found in this relatively simple monatomic system may help to bridge the gap between experiments and theories concerning the universality of amorphous solids.
\begin{acknowledgments}
We thank Dr. Huey-Daw Wu for his help in reactive-ion etching process. We thank Dr. P. M. Ossi for his help during the initial stage of this work. One of the author (X. L.) also greatly appreciated the discussions he had with many experts in the field during the International Workshop on Collective Phenomena in the Low Temperature Physics of Glasses held at the Max-Planck Institut f\"{u}r Physik Komplexer Systeme, Dresden, Germany in 2003. This work was supported by the Office of Naval Research and DARPA.
\end{acknowledgments}

\bibliography{dlc}

\begin{thebibliography}{50}
\expandafter\ifx\csname natexlab\endcsname\relax\def\natexlab#1{#1}\fi
\expandafter\ifx\csname bibnamefont\endcsname\relax
  \def\bibnamefont#1{#1}\fi
\expandafter\ifx\csname bibfnamefont\endcsname\relax
  \def\bibfnamefont#1{#1}\fi
\expandafter\ifx\csname citenamefont\endcsname\relax
  \def\citenamefont#1{#1}\fi
\expandafter\ifx\csname url\endcsname\relax
  \def\url#1{\texttt{#1}}\fi
\expandafter\ifx\csname urlprefix\endcsname\relax\def\urlprefix{URL }\fi
\providecommand{\bibinfo}[2]{#2}
\providecommand{\eprint}[2][]{\url{#2}}

\bibitem[{\citenamefont{Pohl et~al.}(2002)\citenamefont{Pohl, Liu, and
  Thompson}}]{02rmp74-pohl}
\bibinfo{author}{\bibfnamefont{R.~O.} \bibnamefont{Pohl}},
  \bibinfo{author}{\bibfnamefont{X.}~\bibnamefont{Liu}}, \bibnamefont{and}
  \bibinfo{author}{\bibfnamefont{E.}~\bibnamefont{Thompson}},
  \bibinfo{journal}{Rev. Mod. Phys.} \textbf{\bibinfo{volume}{74}},
  \bibinfo{pages}{991} (\bibinfo{year}{2002}).

\bibitem[{\citenamefont{Zeller and Pohl}(1971)}]{71prb4-zeller}
\bibinfo{author}{\bibfnamefont{R.~C.} \bibnamefont{Zeller}} \bibnamefont{and}
  \bibinfo{author}{\bibfnamefont{R.~O.} \bibnamefont{Pohl}},
  \bibinfo{journal}{Phys. Rev. B} \textbf{\bibinfo{volume}{4}},
  \bibinfo{pages}{2029} (\bibinfo{year}{1971}).

\bibitem[{\citenamefont{Hunklinger and Arnold}(1976)}]{76-mason-hunklinger}
\bibinfo{author}{\bibfnamefont{S.}~\bibnamefont{Hunklinger}} \bibnamefont{and}
  \bibinfo{author}{\bibfnamefont{W.}~\bibnamefont{Arnold}}, in
  \emph{\bibinfo{booktitle}{Physical Acoustics XII}}, edited by
  \bibinfo{editor}{\bibfnamefont{W.~D.} \bibnamefont{Mason}} \bibnamefont{and}
  \bibinfo{editor}{\bibfnamefont{R.~N.} \bibnamefont{Thurston}}
  (\bibinfo{publisher}{Academic Press, New York}, \bibinfo{year}{1976}), p.
  \bibinfo{pages}{155}.

\bibitem[{\citenamefont{Anderson et~al.}(1972)\citenamefont{Anderson, Halperin,
  and Varma}}]{72pm25-anderson}
\bibinfo{author}{\bibfnamefont{P.~W.} \bibnamefont{Anderson}},
  \bibinfo{author}{\bibfnamefont{B.~I.} \bibnamefont{Halperin}},
  \bibnamefont{and} \bibinfo{author}{\bibfnamefont{C.~M.} \bibnamefont{Varma}},
  \bibinfo{journal}{Philos. Mag.} \textbf{\bibinfo{volume}{25}},
  \bibinfo{pages}{1} (\bibinfo{year}{1972}).

\bibitem[{\citenamefont{Phillips}(1972)}]{72jltp7-phillips}
\bibinfo{author}{\bibfnamefont{W.~A.} \bibnamefont{Phillips}},
  \bibinfo{journal}{J. Low Temp. Phys.} \textbf{\bibinfo{volume}{7}},
  \bibinfo{pages}{351} (\bibinfo{year}{1972}).

\bibitem[{\citenamefont{J{\"{a}}ckle}(1972)}]{72zpb257-jackle}
\bibinfo{author}{\bibfnamefont{J.}~\bibnamefont{J{\"{a}}ckle}},
  \bibinfo{journal}{Z. Phys. B: Condens. Matter}
  \textbf{\bibinfo{volume}{257}}, \bibinfo{pages}{212} (\bibinfo{year}{1972}).

\bibitem[{\citenamefont{Watson and Pohl}(2003)}]{03prb68-watson}
\bibinfo{author}{\bibfnamefont{S.~K.} \bibnamefont{Watson}} \bibnamefont{and}
  \bibinfo{author}{\bibfnamefont{R.~O.} \bibnamefont{Pohl}},
  \bibinfo{journal}{Phys. Rev. B} \textbf{\bibinfo{volume}{68}},
  \bibinfo{pages}{104203} (\bibinfo{year}{2003}).

\bibitem[{\citenamefont{v.~Haumeder et~al.}(1980)\citenamefont{v.~Haumeder,
  Strom, and Hunklinger}}]{80prl44-haumeder}
\bibinfo{author}{\bibfnamefont{M.}~\bibnamefont{v.~Haumeder}},
  \bibinfo{author}{\bibfnamefont{U.}~\bibnamefont{Strom}}, \bibnamefont{and}
  \bibinfo{author}{\bibfnamefont{S.}~\bibnamefont{Hunklinger}},
  \bibinfo{journal}{Phys. Rev. Lett.} \textbf{\bibinfo{volume}{44}},
  \bibinfo{pages}{84} (\bibinfo{year}{1980}).

\bibitem[{\citenamefont{van~den Berg and
  v.~L{\"{o}}hneysen}(1985)}]{85prl55-vandenberg}
\bibinfo{author}{\bibfnamefont{R.}~\bibnamefont{van~den Berg}}
  \bibnamefont{and}
  \bibinfo{author}{\bibfnamefont{H.}~\bibnamefont{v.~L{\"{o}}hneysen}},
  \bibinfo{journal}{Phys. Rev. Lett.} \textbf{\bibinfo{volume}{55}},
  \bibinfo{pages}{2463} (\bibinfo{year}{1985}).

\bibitem[{\citenamefont{Duquesne and Bellessa}(1985)}]{85pmb52-duquesne}
\bibinfo{author}{\bibfnamefont{J.~Y.} \bibnamefont{Duquesne}} \bibnamefont{and}
  \bibinfo{author}{\bibfnamefont{G.}~\bibnamefont{Bellessa}},
  \bibinfo{journal}{Phil. Mag. B} \textbf{\bibinfo{volume}{52}},
  \bibinfo{pages}{821} (\bibinfo{year}{1985}).

\bibitem[{\citenamefont{Liu et~al.}(1997)\citenamefont{Liu, White{, Jr.}, Pohl,
  Iwanizcko, Jones, Mahan, Nelson, Crandall, and Veprek}}]{97prl78-liu}
\bibinfo{author}{\bibfnamefont{X.}~\bibnamefont{Liu}},
  \bibinfo{author}{\bibfnamefont{B.~E.} \bibnamefont{White{, Jr.}}},
  \bibinfo{author}{\bibfnamefont{R.~O.} \bibnamefont{Pohl}},
  \bibinfo{author}{\bibfnamefont{E.}~\bibnamefont{Iwanizcko}},
  \bibinfo{author}{\bibfnamefont{K.~M.} \bibnamefont{Jones}},
  \bibinfo{author}{\bibfnamefont{A.~H.} \bibnamefont{Mahan}},
  \bibinfo{author}{\bibfnamefont{B.~N.} \bibnamefont{Nelson}},
  \bibinfo{author}{\bibfnamefont{R.~S.} \bibnamefont{Crandall}},
  \bibnamefont{and} \bibinfo{author}{\bibfnamefont{S.}~\bibnamefont{Veprek}},
  \bibinfo{journal}{Phys. Rev. Lett.} \textbf{\bibinfo{volume}{78}},
  \bibinfo{pages}{4418} (\bibinfo{year}{1997}).

\bibitem[{\citenamefont{Liu and Pohl}(1998)}]{98prb58-liu}
\bibinfo{author}{\bibfnamefont{X.}~\bibnamefont{Liu}} \bibnamefont{and}
  \bibinfo{author}{\bibfnamefont{R.~O.} \bibnamefont{Pohl}},
  \bibinfo{journal}{Phys. Rev. B} \textbf{\bibinfo{volume}{58}},
  \bibinfo{pages}{9067} (\bibinfo{year}{1998}).

\bibitem[{\citenamefont{Liu et~al.}(2002)\citenamefont{Liu, Photiadis, Wu,
  Chrisey, Pohl, and Crandall}}]{02pmb82-liu}
\bibinfo{author}{\bibfnamefont{X.}~\bibnamefont{Liu}},
  \bibinfo{author}{\bibfnamefont{D.~M.} \bibnamefont{Photiadis}},
  \bibinfo{author}{\bibfnamefont{H.~D.} \bibnamefont{Wu}},
  \bibinfo{author}{\bibfnamefont{D.~B.} \bibnamefont{Chrisey}},
  \bibinfo{author}{\bibfnamefont{R.~O.} \bibnamefont{Pohl}}, \bibnamefont{and}
  \bibinfo{author}{\bibfnamefont{R.~S.} \bibnamefont{Crandall}},
  \bibinfo{journal}{Phil. Mag. B} \textbf{\bibinfo{volume}{82}},
  \bibinfo{pages}{185} (\bibinfo{year}{2002}).

\bibitem[{\citenamefont{Liu et~al.}(2004)\citenamefont{Liu, Photiadis, Bucaro,
  Vignola, Houston, Wu, and Chrisey}}]{04msea370-liu}
\bibinfo{author}{\bibfnamefont{X.}~\bibnamefont{Liu}},
  \bibinfo{author}{\bibfnamefont{D.~M.} \bibnamefont{Photiadis}},
  \bibinfo{author}{\bibfnamefont{J.~A.} \bibnamefont{Bucaro}},
  \bibinfo{author}{\bibfnamefont{J.~F.} \bibnamefont{Vignola}},
  \bibinfo{author}{\bibfnamefont{B.~H.} \bibnamefont{Houston}},
  \bibinfo{author}{\bibfnamefont{H.~D.} \bibnamefont{Wu}}, \bibnamefont{and}
  \bibinfo{author}{\bibfnamefont{D.~B.} \bibnamefont{Chrisey}},
  \bibinfo{journal}{Mater. Sci. Eng. A} \textbf{\bibinfo{volume}{370}},
  \bibinfo{pages}{142} (\bibinfo{year}{2004}).

\bibitem[{\citenamefont{Phillips}(1979)}]{79jncs34-phillips}
\bibinfo{author}{\bibfnamefont{J.~C.} \bibnamefont{Phillips}},
  \bibinfo{journal}{J. Non-Cryst. Solids} \textbf{\bibinfo{volume}{34}},
  \bibinfo{pages}{153} (\bibinfo{year}{1979}).

\bibitem[{\citenamefont{Thorpe}(1983)}]{83jncs57-thorpe}
\bibinfo{author}{\bibfnamefont{M.~F.} \bibnamefont{Thorpe}},
  \bibinfo{journal}{J. Non-Cryst. Solids} \textbf{\bibinfo{volume}{57}},
  \bibinfo{pages}{355} (\bibinfo{year}{1983}).

\bibitem[{\citenamefont{Brand}(1991)}]{91epl16-brand}
\bibinfo{author}{\bibfnamefont{O.}~\bibnamefont{Brand}},
  \bibinfo{journal}{Europhys. Lett.} \textbf{\bibinfo{volume}{16}},
  \bibinfo{pages}{455} (\bibinfo{year}{1991}).

\bibitem[{rem()}]{remark1}
\bibinfo{note}{Amorphous carbon films with high $sp^3$ concentration prepared
  by various techniques are called diamond-like carbon films, or tetrahedral
  amorphous carbon films in the literature. For simplicity, we use the name
  amorphous carbon films, or {\em a}-C}.

\bibitem[{\citenamefont{Robertson}(2002)}]{02mser37-roberts}
\bibinfo{author}{\bibfnamefont{J.}~\bibnamefont{Robertson}},
  \bibinfo{journal}{Mater. Sci. Eng. R} \textbf{\bibinfo{volume}{37}},
  \bibinfo{pages}{129} (\bibinfo{year}{2002}).

\bibitem[{\citenamefont{Friedmann et~al.}(1997)\citenamefont{Friedmann,
  Sullivan, Knapp, Tallant, Follstaedt, Medlin, and
  Mirkarimi}}]{97apl71-friedmann}
\bibinfo{author}{\bibfnamefont{T.~A.} \bibnamefont{Friedmann}},
  \bibinfo{author}{\bibfnamefont{J.~P.} \bibnamefont{Sullivan}},
  \bibinfo{author}{\bibfnamefont{J.~A.} \bibnamefont{Knapp}},
  \bibinfo{author}{\bibfnamefont{D.~R.} \bibnamefont{Tallant}},
  \bibinfo{author}{\bibfnamefont{D.~M.} \bibnamefont{Follstaedt}},
  \bibinfo{author}{\bibfnamefont{D.~L.} \bibnamefont{Medlin}},
  \bibnamefont{and} \bibinfo{author}{\bibfnamefont{P.~B.}
  \bibnamefont{Mirkarimi}}, \bibinfo{journal}{Appl. Phys. Lett.}
  \textbf{\bibinfo{volume}{71}}, \bibinfo{pages}{3820} (\bibinfo{year}{1997}).

\bibitem[{\citenamefont{Bonelli et~al.}(2000)\citenamefont{Bonelli, Fioravanti,
  Miotello, and Ossi}}]{00epl50-bonelli}
\bibinfo{author}{\bibfnamefont{M.}~\bibnamefont{Bonelli}},
  \bibinfo{author}{\bibfnamefont{A.~P.} \bibnamefont{Fioravanti}},
  \bibinfo{author}{\bibfnamefont{A.}~\bibnamefont{Miotello}}, \bibnamefont{and}
  \bibinfo{author}{\bibfnamefont{P.~M.} \bibnamefont{Ossi}},
  \bibinfo{journal}{Europhys. Lett.} \textbf{\bibinfo{volume}{50}},
  \bibinfo{pages}{501} (\bibinfo{year}{2000}).

\bibitem[{\citenamefont{Bonelli et~al.}(1999)\citenamefont{Bonelli, Cestari,
  and Miotello}}]{99mst10-bonelli}
\bibinfo{author}{\bibfnamefont{M.}~\bibnamefont{Bonelli}},
  \bibinfo{author}{\bibfnamefont{C.}~\bibnamefont{Cestari}}, \bibnamefont{and}
  \bibinfo{author}{\bibfnamefont{A.}~\bibnamefont{Miotello}},
  \bibinfo{journal}{Meas. Sci. Technol.} \textbf{\bibinfo{volume}{10}},
  \bibinfo{pages}{N27} (\bibinfo{year}{1999}).

\bibitem[{\citenamefont{Bonelli et~al.}(2002)\citenamefont{Bonelli, Ferrari,
  Fioravanti, Bassi, Miotello, and Ossi}}]{02epjb25-bonelli}
\bibinfo{author}{\bibfnamefont{M.}~\bibnamefont{Bonelli}},
  \bibinfo{author}{\bibfnamefont{A.~C.} \bibnamefont{Ferrari}},
  \bibinfo{author}{\bibfnamefont{A.~P.} \bibnamefont{Fioravanti}},
  \bibinfo{author}{\bibfnamefont{A.~L.} \bibnamefont{Bassi}},
  \bibinfo{author}{\bibfnamefont{A.}~\bibnamefont{Miotello}}, \bibnamefont{and}
  \bibinfo{author}{\bibfnamefont{P.~M.} \bibnamefont{Ossi}},
  \bibinfo{journal}{Eur. Phys. J. B} \textbf{\bibinfo{volume}{25}},
  \bibinfo{pages}{269} (\bibinfo{year}{2002}).

\bibitem[{\citenamefont{Bonelli et~al.}(2003)\citenamefont{Bonelli, Miotello,
  Mosaner, Casiraghi, and Ossi}}]{03jap93-bonelli}
\bibinfo{author}{\bibfnamefont{M.}~\bibnamefont{Bonelli}},
  \bibinfo{author}{\bibfnamefont{A.}~\bibnamefont{Miotello}},
  \bibinfo{author}{\bibfnamefont{P.}~\bibnamefont{Mosaner}},
  \bibinfo{author}{\bibfnamefont{C.}~\bibnamefont{Casiraghi}},
  \bibnamefont{and} \bibinfo{author}{\bibfnamefont{P.~M.} \bibnamefont{Ossi}},
  \bibinfo{journal}{J. Appl. Phys.} \textbf{\bibinfo{volume}{93}},
  \bibinfo{pages}{859} (\bibinfo{year}{2003}).

\bibitem[{\citenamefont{White{, Jr.} and Pohl}(1995)}]{95mrs356-white}
\bibinfo{author}{\bibfnamefont{B.~E.} \bibnamefont{White{, Jr.}}}
  \bibnamefont{and} \bibinfo{author}{\bibfnamefont{R.~O.} \bibnamefont{Pohl}},
  in \emph{\bibinfo{booktitle}{Thin Films: Stresses and Mechanical Properties
  V}}, edited by \bibinfo{editor}{\bibfnamefont{S.~P.} \bibnamefont{Baker}},
  \bibinfo{editor}{\bibfnamefont{C.~A.} \bibnamefont{Ross}},
  \bibinfo{editor}{\bibfnamefont{P.~H.} \bibnamefont{Townsend}},
  \bibinfo{editor}{\bibfnamefont{C.~A.} \bibnamefont{Volkert}},
  \bibnamefont{and} \bibinfo{editor}{\bibfnamefont{P.}~\bibnamefont{Borgesen}}
  (\bibinfo{year}{1995}), p. \bibinfo{pages}{567}.

\bibitem[{\citenamefont{Li et~al.}(2004)\citenamefont{Li, Fang, Veprek, and
  Li}}]{04msea370-li}
\bibinfo{author}{\bibfnamefont{Z.~S.} \bibnamefont{Li}},
  \bibinfo{author}{\bibfnamefont{Q.~F.} \bibnamefont{Fang}},
  \bibinfo{author}{\bibfnamefont{S.}~\bibnamefont{Veprek}}, \bibnamefont{and}
  \bibinfo{author}{\bibfnamefont{S.~Z.} \bibnamefont{Li}},
  \bibinfo{journal}{Mater. Sci. Eng. A} \textbf{\bibinfo{volume}{370}},
  \bibinfo{pages}{186} (\bibinfo{year}{2004}).

\bibitem[{\citenamefont{Phillips}(1987)}]{87rpp50-phillips}
\bibinfo{author}{\bibfnamefont{W.~A.} \bibnamefont{Phillips}},
  \bibinfo{journal}{Rep. Prog. Phys.} \textbf{\bibinfo{volume}{50}},
  \bibinfo{pages}{1657} (\bibinfo{year}{1987}).

\bibitem[{\citenamefont{Bellessa}(1978)}]{78prl40-bellessa}
\bibinfo{author}{\bibfnamefont{G.}~\bibnamefont{Bellessa}},
  \bibinfo{journal}{Phys. Rev. Lett.} \textbf{\bibinfo{volume}{40}},
  \bibinfo{pages}{1456} (\bibinfo{year}{1978}).

\bibitem[{\citenamefont{White{, Jr.} and Pohl}(1996)}]{96zpb100-white}
\bibinfo{author}{\bibfnamefont{B.~E.} \bibnamefont{White{, Jr.}}}
  \bibnamefont{and} \bibinfo{author}{\bibfnamefont{R.~O.} \bibnamefont{Pohl}},
  \bibinfo{journal}{Z. Phys. B: Condens. Matter}
  \textbf{\bibinfo{volume}{100}}, \bibinfo{pages}{100} (\bibinfo{year}{1996}).

\bibitem[{\citenamefont{Metcalf et~al.}()\citenamefont{Metcalf, Liu, Houston,
  Baldwin, Butler, and Feygelson}}]{metcalf}
\bibinfo{author}{\bibfnamefont{T.~H.} \bibnamefont{Metcalf}},
  \bibinfo{author}{\bibfnamefont{X.}~\bibnamefont{Liu}},
  \bibinfo{author}{\bibfnamefont{B.~H.} \bibnamefont{Houston}},
  \bibinfo{author}{\bibfnamefont{J.~W.} \bibnamefont{Baldwin}},
  \bibinfo{author}{\bibfnamefont{J.~E.} \bibnamefont{Butler}},
  \bibnamefont{and}
  \bibinfo{author}{\bibfnamefont{T.}~\bibnamefont{Feygelson}},
  \bibinfo{note}{submitted to Appl. Phys. Lett.}

\bibitem[{\citenamefont{{Van Cleve}}(1991)}]{vancleve91}
\bibinfo{author}{\bibfnamefont{J.~E.} \bibnamefont{{Van Cleve}}}, Ph.D. thesis,
  \bibinfo{school}{Cornell University} (\bibinfo{year}{1991}),
  \bibinfo{note}{unpublished}.

\bibitem[{\citenamefont{Ferrari et~al.}(1999)\citenamefont{Ferrari, Roberts,
  Geghi, Bottani, Ferulano, and Pastorelli}}]{99apl75-ferrari}
\bibinfo{author}{\bibfnamefont{A.~C.} \bibnamefont{Ferrari}},
  \bibinfo{author}{\bibfnamefont{J.}~\bibnamefont{Roberts}},
  \bibinfo{author}{\bibfnamefont{M.~G.} \bibnamefont{Geghi}},
  \bibinfo{author}{\bibfnamefont{C.~E.} \bibnamefont{Bottani}},
  \bibinfo{author}{\bibfnamefont{R.}~\bibnamefont{Ferulano}}, \bibnamefont{and}
  \bibinfo{author}{\bibfnamefont{R.}~\bibnamefont{Pastorelli}},
  \bibinfo{journal}{Appl. Phys. Lett.} \textbf{\bibinfo{volume}{75}},
  \bibinfo{pages}{1893} (\bibinfo{year}{1999}).

\bibitem[{\citenamefont{Beghi et~al.}(2004)\citenamefont{Beghi, Bottani, Bassi,
  Ossi, Tanner, Ferrari, and Roberts}}]{02msec19-beghi}
\bibinfo{author}{\bibfnamefont{M.~G.} \bibnamefont{Beghi}},
  \bibinfo{author}{\bibfnamefont{C.~E.} \bibnamefont{Bottani}},
  \bibinfo{author}{\bibfnamefont{A.~L.} \bibnamefont{Bassi}},
  \bibinfo{author}{\bibfnamefont{P.~M.} \bibnamefont{Ossi}},
  \bibinfo{author}{\bibfnamefont{B.~K.} \bibnamefont{Tanner}},
  \bibinfo{author}{\bibfnamefont{A.~C.} \bibnamefont{Ferrari}},
  \bibnamefont{and} \bibinfo{author}{\bibfnamefont{J.}~\bibnamefont{Roberts}},
  \bibinfo{journal}{Mater. Sci. Eng. A} \textbf{\bibinfo{volume}{370}},
  \bibinfo{pages}{186} (\bibinfo{year}{2004}).

\bibitem[{\citenamefont{Marques et~al.}(2003)\citenamefont{Marques, Lacerda,
  Champi, Stolojan, Cox, and Silva}}]{03apl83-marques}
\bibinfo{author}{\bibfnamefont{F.~C.} \bibnamefont{Marques}},
  \bibinfo{author}{\bibfnamefont{R.~G.} \bibnamefont{Lacerda}},
  \bibinfo{author}{\bibfnamefont{A.}~\bibnamefont{Champi}},
  \bibinfo{author}{\bibfnamefont{V.}~\bibnamefont{Stolojan}},
  \bibinfo{author}{\bibfnamefont{D.~C.} \bibnamefont{Cox}}, \bibnamefont{and}
  \bibinfo{author}{\bibfnamefont{S.~R.~P.} \bibnamefont{Silva}},
  \bibinfo{journal}{Appl. Phys. Lett.} \textbf{\bibinfo{volume}{83}},
  \bibinfo{pages}{3099} (\bibinfo{year}{2003}).

\bibitem[{\citenamefont{Tanaka}(1989)}]{89prb39-tanaka}
\bibinfo{author}{\bibfnamefont{K.}~\bibnamefont{Tanaka}},
  \bibinfo{journal}{Phys. Rev. B} \textbf{\bibinfo{volume}{39}},
  \bibinfo{pages}{1270} (\bibinfo{year}{1989}).

\bibitem[{\citenamefont{Watson}(1995)}]{95prl75-watson}
\bibinfo{author}{\bibfnamefont{S.~K.} \bibnamefont{Watson}},
  \bibinfo{journal}{Phys. Rev. Lett.} \textbf{\bibinfo{volume}{75}},
  \bibinfo{pages}{1965} (\bibinfo{year}{1995}).

\bibitem[{\citenamefont{Liu et~al.}(1998)\citenamefont{Liu, Vu, Pohl,
  Schiettekatte, and Roorda}}]{98prl81-liu}
\bibinfo{author}{\bibfnamefont{X.}~\bibnamefont{Liu}},
  \bibinfo{author}{\bibfnamefont{P.~D.} \bibnamefont{Vu}},
  \bibinfo{author}{\bibfnamefont{R.~O.} \bibnamefont{Pohl}},
  \bibinfo{author}{\bibfnamefont{F.}~\bibnamefont{Schiettekatte}},
  \bibnamefont{and} \bibinfo{author}{\bibfnamefont{S.}~\bibnamefont{Roorda}},
  \bibinfo{journal}{Phys. Rev. Lett.} \textbf{\bibinfo{volume}{81}},
  \bibinfo{pages}{3171} (\bibinfo{year}{1998}).

\bibitem[{\citenamefont{Blech and Wood}(1993)}]{93jvsta11-blech}
\bibinfo{author}{\bibfnamefont{I.~A.} \bibnamefont{Blech}} \bibnamefont{and}
  \bibinfo{author}{\bibfnamefont{.}~\bibnamefont{Wood}}, \bibinfo{journal}{J.
  Vac. Sci. Technol. A} \textbf{\bibinfo{volume}{11}}, \bibinfo{pages}{728}
  (\bibinfo{year}{1993}).

\bibitem[{\citenamefont{de~Lima{, Jr.} et~al.}(1999)\citenamefont{de~Lima{,
  Jr.}, Lacerda, Vilcarromero, and Marques}}]{99jap86-lima}
\bibinfo{author}{\bibfnamefont{M.~M.} \bibnamefont{de~Lima{, Jr.}}},
  \bibinfo{author}{\bibfnamefont{R.~G.} \bibnamefont{Lacerda}},
  \bibinfo{author}{\bibfnamefont{J.}~\bibnamefont{Vilcarromero}},
  \bibnamefont{and} \bibinfo{author}{\bibfnamefont{F.~C.}
  \bibnamefont{Marques}}, \bibinfo{journal}{J. Appl. Phys.}
  \textbf{\bibinfo{volume}{86}}, \bibinfo{pages}{4936} (\bibinfo{year}{1999}).

\bibitem[{\citenamefont{Dyrting et~al.}(1999)\citenamefont{Dyrting, Fang,
  Szeto, and Sheng}}]{99ep48-dyrting}
\bibinfo{author}{\bibfnamefont{S.}~\bibnamefont{Dyrting}},
  \bibinfo{author}{\bibfnamefont{H.~P.} \bibnamefont{Fang}},
  \bibinfo{author}{\bibfnamefont{K.~Y.} \bibnamefont{Szeto}}, \bibnamefont{and}
  \bibinfo{author}{\bibfnamefont{P.}~\bibnamefont{Sheng}},
  \bibinfo{journal}{Europhys. Lett.} \textbf{\bibinfo{volume}{48}},
  \bibinfo{pages}{403} (\bibinfo{year}{1999}).

\bibitem[{\citenamefont{Rivier and Wooten}(2003)}]{03cmcc48-rivier}
\bibinfo{author}{\bibfnamefont{N.}~\bibnamefont{Rivier}} \bibnamefont{and}
  \bibinfo{author}{\bibfnamefont{F.}~\bibnamefont{Wooten}},
  \bibinfo{journal}{Commun. in Math. and in Comp. Chem.}
  \textbf{\bibinfo{volume}{48}}, \bibinfo{pages}{145} (\bibinfo{year}{2003}).

\bibitem[{\citenamefont{Klein et~al.}(1978)\citenamefont{Klein, Fischer,
  Anderson, and Anthony}}]{78prb15-klein}
\bibinfo{author}{\bibfnamefont{M.~W.} \bibnamefont{Klein}},
  \bibinfo{author}{\bibfnamefont{B.}~\bibnamefont{Fischer}},
  \bibinfo{author}{\bibfnamefont{A.~C.} \bibnamefont{Anderson}},
  \bibnamefont{and} \bibinfo{author}{\bibfnamefont{P.~J.}
  \bibnamefont{Anthony}}, \bibinfo{journal}{Phys. Rev. B}
  \textbf{\bibinfo{volume}{15}}, \bibinfo{pages}{5887} (\bibinfo{year}{1978}).

\bibitem[{\citenamefont{Burin et~al.}(1998)\citenamefont{Burin, Natelson,
  Osheroff, and Kagan}}]{burin98}
\bibinfo{author}{\bibfnamefont{A.~L.} \bibnamefont{Burin}},
  \bibinfo{author}{\bibfnamefont{D.}~\bibnamefont{Natelson}},
  \bibinfo{author}{\bibfnamefont{D.~D.} \bibnamefont{Osheroff}},
  \bibnamefont{and} \bibinfo{author}{\bibfnamefont{Y.}~\bibnamefont{Kagan}}, in
  \emph{\bibinfo{booktitle}{Tunneling Systems in Amorphous and Crystalline
  Solids}}, edited by
  \bibinfo{editor}{\bibfnamefont{P.}~\bibnamefont{Esquinazi}}
  (\bibinfo{publisher}{Springer, Berlin}, \bibinfo{year}{1998}), p.
  \bibinfo{pages}{223}.

\bibitem[{\citenamefont{Strehlow et~al.}(1998)\citenamefont{Strehlow, C.Enss,
  and Hunklinger}}]{98prl80-strehlow}
\bibinfo{author}{\bibfnamefont{P.}~\bibnamefont{Strehlow}},
  \bibinfo{author}{\bibnamefont{C.Enss}}, \bibnamefont{and}
  \bibinfo{author}{\bibfnamefont{S.}~\bibnamefont{Hunklinger}},
  \bibinfo{journal}{Phys. Rev. Lett.} \textbf{\bibinfo{volume}{80}},
  \bibinfo{pages}{5361} (\bibinfo{year}{1998}).

\bibitem[{\citenamefont{K{\"{o}}nig et~al.}(2002)\citenamefont{K{\"{o}}nig,
  Ramos, Usherov-Marshak, Arcas-Guijarro, Hernando-Ma{\~{n}}eru, and
  Esquinazi}}]{02prb65-konig}
\bibinfo{author}{\bibfnamefont{R.}~\bibnamefont{K{\"{o}}nig}},
  \bibinfo{author}{\bibfnamefont{M.~A.} \bibnamefont{Ramos}},
  \bibinfo{author}{\bibfnamefont{I.}~\bibnamefont{Usherov-Marshak}},
  \bibinfo{author}{\bibfnamefont{J.}~\bibnamefont{Arcas-Guijarro}},
  \bibinfo{author}{\bibfnamefont{A.}~\bibnamefont{Hernando-Ma{\~{n}}eru}},
  \bibnamefont{and}
  \bibinfo{author}{\bibfnamefont{P.}~\bibnamefont{Esquinazi}},
  \bibinfo{journal}{Phys. Rev. B} \textbf{\bibinfo{volume}{65}},
  \bibinfo{pages}{180201} (\bibinfo{year}{2002}).

\bibitem[{\citenamefont{Ladieu et~al.}(2003)\citenamefont{Ladieu, Cochec, Pari,
  Trouslard, and Ailloud}}]{03prl90-ladieu}
\bibinfo{author}{\bibfnamefont{F.}~\bibnamefont{Ladieu}},
  \bibinfo{author}{\bibfnamefont{J.~L.} \bibnamefont{Cochec}},
  \bibinfo{author}{\bibfnamefont{P.}~\bibnamefont{Pari}},
  \bibinfo{author}{\bibfnamefont{P.}~\bibnamefont{Trouslard}},
  \bibnamefont{and} \bibinfo{author}{\bibfnamefont{P.}~\bibnamefont{Ailloud}},
  \bibinfo{journal}{Phys. Rev. Lett.} \textbf{\bibinfo{volume}{90}},
  \bibinfo{pages}{205501} (\bibinfo{year}{2003}).

\bibitem[{\citenamefont{Classen et~al.}(2000)\citenamefont{Classen, Burkert,
  Enss, and Hunklinger}}]{00prl84-classen}
\bibinfo{author}{\bibfnamefont{J.}~\bibnamefont{Classen}},
  \bibinfo{author}{\bibfnamefont{T.}~\bibnamefont{Burkert}},
  \bibinfo{author}{\bibfnamefont{C.}~\bibnamefont{Enss}}, \bibnamefont{and}
  \bibinfo{author}{\bibfnamefont{S.}~\bibnamefont{Hunklinger}},
  \bibinfo{journal}{Phys. Rev. Lett.} \textbf{\bibinfo{volume}{84}},
  \bibinfo{pages}{2176} (\bibinfo{year}{2000}).

\bibitem[{\citenamefont{Parshin}(1994)}]{94pss36-parshin}
\bibinfo{author}{\bibfnamefont{D.~A.} \bibnamefont{Parshin}},
  \bibinfo{journal}{Phys. Solid State} \textbf{\bibinfo{volume}{36}},
  \bibinfo{pages}{991} (\bibinfo{year}{1994}).

\bibitem[{\citenamefont{K{\"{u}}hn}(20023)}]{03epl62-kuhn}
\bibinfo{author}{\bibfnamefont{R.}~\bibnamefont{K{\"{u}}hn}},
  \bibinfo{journal}{Europhys. Lett.} \textbf{\bibinfo{volume}{652}},
  \bibinfo{pages}{313} (\bibinfo{year}{20023}).

\bibitem[{\citenamefont{Turlakov}(2004)}]{04prl93-turlakov}
\bibinfo{author}{\bibfnamefont{M.}~\bibnamefont{Turlakov}},
  \bibinfo{journal}{Phys. Rev. Lett.} \textbf{\bibinfo{volume}{93}},
  \bibinfo{pages}{035501} (\bibinfo{year}{2004}).

\end{thebibliography}

\end{document}